 \documentstyle[12pt]{article}
\begin{document}
 \newcommand{\be}[1]{\begin{equation}\label{#1}}
 \newcommand{\ee}{\end{equation}}
 \newcommand{\beqn}[1]{\begin{eqnarray}\label{#1}}
 \newcommand{\eeqn}{\end{eqnarray}}
\newcommand{\mat}[4]{\left(\begin{array}{cc}{#1}&{#2}\\{#3}&{#4}\end{array}
\right)}
 \newcommand{\matr}[9]{\left(\begin{array}{ccc}{#1}&{#2}&{#3}\\{#4}&{#5}&{#6}\\
{#7}&{#8}&{#9}\end{array}\right)}
 \newcommand{\eps}{\varepsilon}
 \newcommand{\Ga}{\Gamma}
 \newcommand{\la}{\lambda}
\newcommand{\ov}{\overline}
\newcommand{\mucirc}{\stackrel{\circ}{\mu}}
\newcommand{\mast}{\stackrel{\ast}{m}}
\newcommand{\meps}{\stackrel{\circ}{\eps}}
\newcommand{\mcirc}{\stackrel{\circ}{m}}
\newcommand{\mcir}{\stackrel{\circ}{M}}
\newcommand{\geqsim}{\stackrel{>}{\sim}}
\renewcommand{\thefootnote}{\fnsymbol{footnote}}

\begin{titlepage}
\begin{flushright}
hep-ph/9511221 \\
INFN-FE 16-95 \\
UMD-PP-96-15\\
IFIC/95-64\\
FTUV/95-61\\
October 1995
\end{flushright}
\vspace{10mm}

 \begin{center}
 {\Large \bf Asymmetric Inflationary Reheating  \\ [2mm]
             and the Nature of Mirror Universe}\\

\vspace{1.3cm}
{\large  Z.G. Berezhiani$^{a,b}$,
{}~A.D. Dolgov$^{c,d}$
{}~and ~ R.N. Mohapatra$^e$ }
\\ [5mm]
$^a$ {\em Istituto Nazionale di Fisica Nucleare, Sezione di Ferrara,
44100 Ferrara, Italy \\ [2mm]
$^b$ Institute of Physics, Georgian Academy
of Sciences, 380077 Tbilisi, Georgia}\\ [2mm]
$^c$ {\em I.F.I.C.-C.S.I.C., Universitad de Valencia, 46100 Burjassot,
Valencia, Spain } \\ [2mm]
$^d$ {\em ITEP, 117259 Moscow, Russia } \\ [2mm]
$^e$ {\em Department of Physics, University of Maryland, College Park,
MD 20742, U.S.A.}
\end{center}

\vspace{2mm}
\begin{abstract}

The existence of a shadow world (or mirror universe) with matter and
forces identical to that of the visible world  but interacting with
the latter only via gravity can be motivated by superstring theories
as well as by recent attempts to understand the nature of a sterile
neutrino needed if all known neutrino data are to be consistent with
each other. A simple way to reconcile the constraints of big bang
nucleosynthesis in such a theory is to postulate that the reheating
temperature after inflation in the mirror universe is lower than that
in the visible one. We have constructed explicit models that realize
this proposal and have shown that the asymmetric reheating can be
related to a difference of the electroweak symmetry breaking scales
in the two sectors, which is needed for a solution of the neutrino
puzzles in this picture. Cosmological implications of the mirror
matter are also discussed.

\end{abstract}

\end{titlepage}

\renewcommand{\thefootnote}{\arabic{footnote})}
\setcounter{footnote}{0}

\noindent
{\large \bf 1. Introduction}
\vspace{4mm}

The $E_8\times E^{\prime}_8$ string theories indicate an
interesting possibility that microphysics of the early universe has
two parallel components with identical matter and force structure
which communicate only through gravity \cite{Witten}.
In a recent paper \cite{BM}, two of us have suggested that the same
idea may be motivated by the existing neutrino data
(for a different model see also \cite{FV}).\footnote{The concept of
a hidden mirror world has been considered also in several earlier
papers \cite{mirror}. }
The reasoning goes as follows: the simplest way to reconcile
the present neutrino puzzles is to invoke a fourth neutrino which
is sterile with respect to the weak interactions and extremely
light (with the mass in the milli-eV range) \cite{calmoh}.
The solar neutrino problem (SNP) can be
then explained by the MSW mechanism operating between the $\nu_e$
and $\nu_s$ whereas the atmospheric neutrino data is
explained by the oscillation between the nearly degenerate
$\nu_{\mu}$ and $\nu_{\tau}$ states with mass of 2.5 eV
which also provide a cosmological hot dark matter (HDM).
The recent LSND results can be also explained by small oscillations
between $\nu_e$ and $\nu_\mu$.
{}From a theoretical point of view, the lightness of the sterile neutrino
would be easier to understand if it could be subjected
to the same kind of symmetry reasoning that keeps the known neutrinos
light, i.e. accidental B-L symmetry possibly broken by gravity or
a local B-L symmetry broken at some very high scale as in the usual
seesaw mechanism. If one postulates a mirror universe with identical
gauge and matter content \cite{mirror},
the neutrinos of the mirror universe become
subjected to the mirror B-L symmetry and remain light.
In particular, the sterile state $\nu_s$ can be the mirror partner
$\nu'_e$ of the usual electron neutrino.
In fact it is shown in Ref. \cite{BM} that using only one input that
the electroweak symmetry breaking scale in the mirror universe
is a factor $\zeta\sim 30$ higher than the usual electroweak scale,
simultaneously gives both the desired mass and mixing range for the
 MSW oscillation $\nu_e-\nu'_e$ to be successful
in solving the solar neutrino problem.
In addition, if $\nu_{\mu}$ and $\nu_\tau$ have masses in eV range
constituting thereby HDM, their mirror partners $\nu'_\mu$ and
$\nu'_\tau$ being factor of $\zeta^2\sim 1000$ heavier,
will have masses in keV range
and thus can provide the warm dark matter.

An immediate challenge to this approach is to reconcile the constraints
of big bang nucleosynthesis (BBN) on the energy density of the universe at
the BBN epoch, which translates into constraints on the number of light
neutrinos $N_{\nu}$ \cite{sch}. In a theory such as ours, since all mirror
neutrinos (and photons) are light, they could apriori give a large
contribution considerably exceeding $N_{\nu}= 3$.
Therefore, for our idea to be viable,
the contribution of the light mirror particles to the energy density
at the BBN epoch must be appropriately reduced.
The idea to achieve this goal suggested in Ref. \cite{BM}
was to postulate an asymmetric postinflationary reheating of the two
universes. In particular, if the mirror universe reheats to a lower
temperature than our universe, then BBN constraint can be satisfied.
The purpose of the present paper is to present realizations
of this idea in the context of gauge models and then study cosmological
consequences of this hypothesis. We also discuss the state of the mirror
universe at the present epoch. In particular, we argue that for the case
when the electroweak symmetry breaking in the mirror sector is about
30 times larger than that of the visible universe, it is likely that
the mirror baryonic matter would consist entirely of
the mirror hydrogen which might be the only stable mirror atom.
Its implications for the microlensing experiments are also discussed.

%

\vspace{8mm}
\noindent
{\large \bf 2. Mirror world: a brief review and cosmological implications}
\vspace{4mm}

Having in mind the $E_8\times E_8'$ string model,
one can imagine that below the Planck (string) scale field theory
is given by a product of two identical gauge groups $G\times G'$
with identical particle contents, and there is
a discrete symmetry $P(G\leftrightarrow G')$ interchanging
all particles in corresponding representations of $G$ (which we
consider as visible world) and $G'$ (which we call mirror world).
It guarantees that all bare coupling constants
(gauge, Yukawa, Higgs) have the same values in both sectors.
We also assume that the two worlds communicate only through
gravity and possibly also via a heavy gauge singlet matter field.
At some scale below $M_{Pl}$ the gauge symmetry breaks down
to $G_{\rm SM}\times G'_{\rm SM}$,
where $G_{\rm SM}=SU(3)\times  SU(2)\times U(1)$ stands for the
standard model incorporating quarks and leptons
$q_i, ~u^c_i, ~d^c_i; ~l_i, ~e^c_i$ and Higgs doublet $\phi$,
and $G'_{\rm SM}=[SU(3)\times SU(2)\times U(1)]'$ is its mirror
counterpart with analogous particle content:
$q'_i, ~u'^c_i, ~d'^c_i; ~l'_i, ~e'^c_i$  and $\phi'$
($i=1,2,3$ is a family index).\footnote{It is essential
that at higher energies both $SU(3)\times SU(2)\times U(1)$ gauge factors
are embedded into simple gauge groups.
Otherwise kinetic terms of the two $U(1)$ gauge fields could mix and
this would impart arbitrary electric charges to the particles
\cite{mirror}. In the spirit of this proposal, one may therefore
envisage that the gauge groups $G$ and $G'$ are identical and simple
(e.g. $SU(5)$, $SU(6)$, $SO(10)$ or any other GUT subgroup of $E_8$).
Even in this case, the kinetic mixing between usual and mirror photons
would arise from radiative effects in the presence of the
mixed representations of $G\times G'$. Bearing in mind the possible
$E_8\times E_8'$ string origin for our models, here
we exclude such mixed representations.}
$P$ parity remains unbroken at this stage, so that
the coupling constants of the two sectors evolve down in energy
from common values. Let us also assume that
there exists a mechanism that spontaneously breaks $P$ parity
at lower energies and thus allows the two electroweak scales
$\langle \phi' \rangle =v'$ and $\langle \phi \rangle =v$
to be different: we assume that $v'\gg v$.
As far as the Yukawa couplings have the same values in both systems,
the mass and mixing pattern of the charged fermions in the mirror world
is completely analogous to that of the visible one, but with all
fermion masses scaled up by the factor $\zeta=v'/v$.
The masses of gauge bosons and higgses are also scaled as
$M_{W',Z',\phi'}=\zeta M_{W,Z,\phi}$
while photons and gluons remain massless in both sectors.

As for the neutrino masses, they can emerge only via operators
bilinear in the Higgs fields and cutoff by a large scale $M$,
which can be effectively induced for example via the seesaw
mechanism.
On general grounds, by assuming that $P$ parity breaks at lower
energies ($E\ll M$), these operators can be written as
\be{LRnu}
\frac{h_{ij}}{M} (l_i\phi)C(l_j\phi) +
\frac{h_{ij}}{M} (l'_i\phi')C(l'_j\phi') + {\rm h.c.}
\ee
where $C$ is a charge conjugation matrix.
 In order to deal with the present neutrino data, one can assume
further that $M\sim 10^{13}$ GeV and that
the $O(1)$ coupling constants $h_{ij}$
obey an approximate $ L_e+L_\mu-L_\tau$ symmetry \cite{BM}.
Thus $\nu_e$ and $\nu'_e$ are left massless,
$\nu_{\mu,\tau}$ get almost degenerate masses $m\sim v^2/M\sim$ few eV,
and the masses of their mirror partners $\nu'_{\mu,\tau}$ are
$m'=\zeta^2 m$. The $\nu_e$ and $\nu'_e$ states then can get masses
through the gravity induced Planck scale effects \cite{BEG,ABS}
which explicitly violate the global lepton number, and also mix
the neutrino states of two sectors \cite{ABS}.
The relevant operators are:
\be{Planck}
\frac{\alpha_{ij}}{M_{Pl}} (l_{i}\phi)C(l_{j}\phi) +
\frac{\alpha_{ij}}{M_{Pl}} (l'_{i}\phi')C(l'_{j} \phi')
+ \frac{\beta_{ij}}{M_{Pl}} (l_i\phi)C(l'_{j} \phi') + {\rm h.c.}
\ee

\noindent
with constants $\alpha,\beta\sim 1$.
Then $\nu_e$ and $\nu'_e$ acquire masses respectively
$\sim\ \mcirc$ and  $\sim \zeta^2\!\mcirc $,
while their mixing term is  $\sim \zeta\mcirc $,
where $\mcirc\ =v^2/M_{Pl}= 2.5\cdot 10^{-6}~ \mbox{eV}$.
Hence, the $\nu_e-\nu'_e$ oscillation emerges with parameters in the
range
\be{alter}
\delta m^2\sim (\zeta/30)^4 \times 6\cdot 10^{-6}\,\mbox{eV}^2, ~~~~~
\sin^2 2\theta\sim
(30/\zeta)^2\times 5\cdot 10^{-3}
\ee
which for $\zeta\sim 30$ perfectly fit the
``small mixing angle'' MSW solution to the SNP \cite{MSW}.
More generally, by taking into account the solar model uncertainties,
as well as possible order of magnitude spread
in the constants $\alpha,\beta$, the relevant range for $\zeta$
can be extended to $\zeta\sim 10-100$ \cite{BM}.
Alternatively,
for $\zeta\sim 1$ we get $\delta m^2 \sim 10^{-11}\,\mbox{eV}^2$
and $\sin^2 2\theta \sim 1$, which corresponds to another solution
for SNP known as `just-so' scenario \cite{justso}.
Independently of the value of $\zeta$, the values of $\delta m^2$
and $\sin^2 2\theta$ given by eq. (\ref{alter})
are safely below the BBN bounds on the
active to sterile neutrino oscillation $\nu_e -\nu'_e$
\cite{bd} even if a very strong upper bound
$\Delta N_\nu < 0.1$ is taken. The same is true for the oscillations
$\nu_{\mu,\tau} - \nu'_{\mu,\tau}$, with mixing between
them induced by the Planck scale operators (\ref{Planck}).
However, the oscillations between $\nu_{\mu,\tau}$ and $\nu'_e$
with $\delta m^2\simeq -m^2$ and
$\sin^2 2\theta\simeq 4\zeta^2 (\mcirc/m)^2$ may possess a resonant
behaviour in the cosmic plasma, and in accordance with the estimates
of \cite{bd} we roughly get an upper limit $\zeta < 10^3$. This
excludes very high values of the mirror electroweak scale and thus
supports our proposal that the $P$ parity breaking is a lower energy
phenomenon.

%

With regard to the two chromodynamics,
a big difference between the electroweak scales $v'$ and $v$
will not cause the similar big
difference between the confinement scales
in two worlds. Indeed, if $P$ parity is valid at higher (GUT) scales,
the strong coupling constants in both sectors would evolve down in energy
with same values until the energy
reaches the value of the mirror-top ($t'$) mass.
Below it $\alpha'_{s}$ will have a different slope than $\alpha_{s}$.
It is then very easy to calculate the value of the scale $\Lambda'$
at which $\alpha'_{s}$ becomes large.
This value of course depends on the ratio $\zeta= v'/v$.
Taking $\Lambda=200$ MeV for the ordinary QCD,
we find $\Lambda'\simeq 280$ MeV if $\zeta\sim 30$.
On the other hand, we have $m'_{u,d}=\zeta m_{u,d}\sim m_s$ so that
masses of the mirror light quarks $u'$ and $d'$
do not exceed $\Lambda'$. So the condensates
$\langle \bar{q}' q' \rangle$ should be formed with
approximately the same magnitudes as
the usual quark condensates $\langle \bar{q}q \rangle$.
As a result, mirror pions should have mass
$m'_{\pi}\simeq \sqrt{m_{u',d'}\langle \bar q' q' \rangle }$
comparable to the mass of normal kaons
$m_{K}\simeq \sqrt{m_{s}\langle \bar q q \rangle }$.

As for the mirror nucleons,
their masses are approximately 1.5 times larger
than that of the usual nucleons.
Since $(m'_d-m'_u)\approx 30 (m_d-m_u)$
we expect the mirror neutron $n'$ to be heavier than the mirror
proton $p'$ by about 150 MeV or so,
while the mirror electron mass is $m'_e=\zeta m_e\sim 15$ MeV.
Clearly, such a large mass difference cannot be compensated by
the nuclear binding energy and hence
even bound neutrons will be unstable
against $\beta$ decay $n'\to p' e' \bar{\nu}'_e$.
Thus in the mirror world
hydrogen will be the only stable nucleus.

Concerning thermodynamics of the two worlds in the Early Universe,
we assume that already at the postinflationary ``reheating'' stage
they are decoupled from each other. As we discuss in next section,
once $P$-invariance is spontaneously broken, it can be violated
also in the inflaton couplings to matter.
Then the inflaton should decay into visible and mirror
particles with different rates, so that after inflation
the temperatures of the ordinary ($T_R$)
and mirror ($T'_R$) thermal bathes would be
different (the idea of using inflation to provide a temperature
difference between ordinary matter and mirror or other forms of
hidden matter was first discussed in ref. \cite{KST}). In this way,
the present cosmological abundance of light mirror particles
can be suppressed as compared to that of their visible partners.

In the standard cosmology the effective number
of the light degrees of freedom
at the BBN era is $g_\ast = 10.75$
as is contributed by photons $\gamma$, $e^+ e^-$ pairs, and
three neutrino species $\nu_{e,\mu,\tau}$, in
a good agreement with the observed light element abundances
\cite{sch}. In the presence of the mirror universe
mirror photons $\gamma'$ and neutrinos $\nu'_{e,\mu,\tau}$
would also contribute the effective number of neutrino
species $N_\nu$:
\be{BBN}
\Delta g_\ast =1.75\,\Delta N_\nu =
(2 + 5.25 x^4) \left(\frac{T'}{T}\right)^4 ,
{}~~~~~~~~ x=T'_\nu/T'
\ee
Here $T$, $T'$ and $T'_\nu$ are respectively the temperatures of
$\gamma$, $\gamma'$ and $\nu'$ at the BBN era.
The value of $x$ is determined by the
temperature $T'_D$ at which $\nu'$s decouple from the mirror
electromagnetic plasma.
$T'_D$ can be expressed through the decoupling temperature
of the ordinary neutrinos ($T_{D(\nu_e)} \simeq 2$ MeV
and $T_{D(\nu_\mu)} \simeq 3$ MeV).
It is scaled as
$T'_D \sim \zeta^{4/3} (g_*T^4/10.75 T'^4 + g'_* /10.75) ^{1/6} T_D$,
where the first factor is related to the larger masses of the mirror
intermediate bosons and the second one comes because the universe
expansion is dominated by the ordinary particles.
If $\zeta \simeq 30$ and $T>2T'$, then mirror neutrinos
decouple before the mirror QCD phase transition:
$T'_D > \Lambda'\simeq 280~\mbox{MeV} \geq m'_{u,d}=\zeta m_{u,d}$,
so that the light quarks $u',d'$ and mirror gluons also contribute
along with the electron $e'$ to the
heating of $\gamma'$. This leads to $x=(4/85)^{1/3}$
and by taking $\Delta N_\nu < 0.1$
we obtain\footnote{One can easily check that
this value of $x$ remains constant up to $\zeta\sim 10^5$, and
then decreases step-by-step due to contributions of the heavier states
$\mu',s'$ etc. On the contrary, for $\zeta\leq 10$ the decoupling of
$\nu'$ occurs below $\Lambda'$ and we
arrive to the `standard' result $x=T'_\nu/T'=(4/11)^{1/3}$. }
\be{T'/T}
\frac{T'}{T}\approx 0.95\, (\Delta N_\nu)^{1/4}
< 0.52
\ee
According to this bound the present day abundance of the mirror
neutrinos relative to the usual ones $r=n_{\nu'}/n_\nu= (xT'/T)^3$
should be less than $10^{-2}$. The usual and mirror neutrinos
contribute to the present cosmological density as
\be{HWDM}
\sum m_\nu = \Omega_{\nu} \cdot 94 h^2 ~\mbox{eV},~~~~~
\sum m'_\nu =
\zeta^2 \sum m_\nu = r^{-1} \Omega'_{\nu} \cdot 94 h^2 ~\mbox{eV}
\ee
where $h$ is the Hubble constant in units 100 Km s$^{-1}$ Mpc$^{-1}$.
One can assume further that ordinary neutrinos have mass
in the eV range and thus form the HDM component
with $\Omega_\nu\simeq 0.2$
(as in the model \cite{BM} with almost degenerate
$\nu_\mu$ and $\nu_\tau$ having masses of about $2.5$ eV).
Then the mirror neutrino masses being factor of $\zeta^2$ larger
emerge in the keV range and thus could constitute the
warm dark matter (WDM) of the universe.
{}From (\ref{HWDM}) we get $r=\zeta^{-2}(\Omega'_\nu/\Omega_\nu)$.
Then taking a rather conservative bound
$\Omega'_{\nu}<0.8$ (bearing in mind that other particles like LSP
or mirror baryons could also contribute to the present energy density),
we obtain the upper bound  comparable to that of Eq. (\ref{T'/T}):
\be{WDM}
\frac{T'}{T} < \frac{1.6}{x \zeta^{2/3}} \approx
0.4\, \left(\frac{30}{\zeta}\right)^{\frac{2}{3}}
\ee

The obtained limits on $T'/T$ can be immediately translated to the
limit on the ratio of the postinflationary reheating temperatures.
Indeed, the two worlds are decoupled and if
during the universe expansion both of them evolve adiabatically with
separately conserved entropy
then we arrive to the nucleosynthesis epoch having
\footnote{We also assume that initially $g_* = g'_*$ despite different
$T_R$ and $T'_R$, which is natural if $T_R,T'_R \gg v'$.
In addition, the relation (\ref{T_R}) holds if there are no first order
phase transitions. In the presence of the latter the situation
could be very much different (see e.g. model of Section 4). }
\be{T_R}
\frac{T'_R}{T_R}=
\left(\frac{2+5.25x^3}{10.75}\right)^{\frac{1}{3}} \frac{T'}{T}\approx
0.6\, \frac{T'}{T}
\ee

Hence, in the most interesting case the electroweak scale $v'$
in the mirror sector should be by factor
$\zeta\sim 30$ larger than the standard electroweak scale
$v=174$ GeV, while the
reheating temperature of the mirror universe
should be 4-5 times smaller than that of the visible one.
In this case
cosmological dark matter can consist dominantly of neutrinos.
In particular, if the ordinary neutrinos $\nu_{\mu,\tau}$ with the
mass of few eV's form the HDM component with $\Omega_{\nu}\simeq 0.2$,
then their mirror partners $\nu'_{\mu,\tau}$ being
$\zeta^2\sim 1000$ times heavier emerge in keV range and with the
present abundance smaller by two orders of magnitude than that of
normal neutrinos, would constitute the WDM with $\Omega'_\nu\simeq 0.7$.
Clearly, their masses satisfy the Tremaine-Gunn limit \cite{TG}
and thus could constitute dark matter even in dwarf spheroidal
galaxies where this limit is most stringent ($m'_\nu> 0.3-0.5$ keV).
The implications of such mixed HDM+WDM scenario for the large scale
structure are rather similar \cite{WDM}
to that of the currently popular HDM+CDM scheme \cite{HDM2}
with the cold dark matter (CDM) consisting of
heavy ($m\sim 100$ GeV) particles or axionic condensate.
However, more detailed observational data
on the large scale structure of the universe may make it possible to
discriminate between warm and cold dark matter.
Moreover, dark matter consisting of sterile neutrinos invalidates
direct searches of the CDM candidates via superconducting detectors
or axion haloscopes. High energy neutrino fluxes from
the Sun and from the Galactic center
which are expected from the annihilation of LSP's if they dominate
dark matter in the universe, will also be absent.
One should however keep in mind that in supersymmetric versions
of our scheme CDM as well could exist in the form of the LSP.

An interesting question is what is the amount and form of
the mirror baryonic dark matter in the universe.
Most likely, baryogenesis in the mirror
universe proceeded through the same mechanism as in the visible one
and we may expect that the baryon
asymmetries (BA) in both worlds should be nearly the same.
Since mirror nucleons are not much heavier than the usual ones
their fraction in the present energy density,
$\Omega_{B'}$, would be about the same as $\Omega_B$,
that is around a few percent.

Let us discuss now cosmological evolution of mirror baryonic
matter. Since the binding energy of the mirror hydrogen atom
is thirty times larger than that of the ordinary hydrogen, its
recombination occurs much earlier than the usual recombination era.
Hence, the evolution of density fluctuations in the mirror matter
would be more efficient than in the visible one.
(From the viewpoint of the visible observer
mirror baryons behave as a dissipative dark matter.)
As a result, one can expect
that the distribution of mirror baryons in galactic discs
should be more clumped towards the center. It is noteworthy that
mirror dark matter may show antibiasing behaviour ($b<1$) which
is considered unphysical for normal dark matter.
Since mirror hydrogen is the only stable nucleus in the mirror world,
nuclear burning could not be ignited and
luminous (in terms of $\gamma'$) mirror stars cannot be formed.
Therefore, nothing can prevent the sufficiently big protostars from
gravitational collapse and in dense regions like galactic cores a
noticeable fraction of mirror baryons would collapse into black holes.
Recent observational data indeed suggest a presence of giant black
holes with masses $\sim 10^{6-7} M_\odot$ in galactic centers.
In addition, easier formation of mirror black holes may explain the
early origin of quasars.

The remaining fraction of the mirror baryons
could fragment into smaller objects like  white dwarves
(or possibly neutron stars) which can maintain stability
due to the pressure of degenerate fermions.
For the mirror stars consisting entirely of hydrogen,
the Chandrasekhar limit is
$M'_{\rm Ch}\simeq 5.75 (m_p/m'_p)^2 M_\odot$,
where $m_p$ and $m'_p\simeq 1.5 m_p$
are respectively the masses of usual and mirror proton.
For smaller mirror objects the evaporation limit
should be $2-3$ orders of magnitude smaller than for the
visible ones because the Bohr radius of the mirror hydrogen is
30 times smaller than that of the usual one.

These mirror objects, being dark for the normal observer,
could be observed as Machos in the gravitational microlensing
experiments (for a review, see e.g. \cite{Ansari}).
In principle they can be distinguished from
the Machos of the visible world.
The latter presumably consist of the dim compact
objects (brown dwarves) too light to burn hydrogen, with
masses ranging from the evaporation limit
$(\sim 10^{-7} M_\odot)$ to the ignition limit
$(\sim 10^{-1} M_\odot)$ \cite{DeRuj}.
The mass spectrum of mirror Machos
extends from the evaporation limit $\sim 10^{-9} M_\odot$
up to the Chandrasekhar limit $M'_{\rm Ch}\simeq 3M_\odot$.
The present data on the microlensing events
are too poor to allow any conclusion on the presence of such
heavy (or light) objects. An unambiguous determination of the Macho
mass for each event is impossible, and only the most probable mass
can be obtained, depending of the spatial and velocity distribution
of Machos. The optical depth or the fraction of the sky covered
by the Einstein disks of Machos,  is nearly
independent of their mass: the Einstein disk surface
is proportional to $M$, while the number of deflectors for a given
total mass decreases as $M^{-1}$.
However, larger event statistics will allow to find
the Macho mass distribution with a better precision.

As noted earlier, the distribution of mirror baryonic matter
in galaxies should be more shifted towards their centers as compared
to the visible matter. Thus one can expect that
mirror stars in our Galaxy would significantly contribute to the
microlensing events towards the galactic bulge, while their
contribution to the microlensing in halo should be smaller than that
of usual Machos.
Interestingly, the event rates in the galactic
bulge observed by OGLE and MACHO experiments are about twice
larger than the expected value deduced from the low mass star population
in the Galactic disk \cite{OGLE}.
Barring accidental conspiracies like a
presence of bar (elongated dense stellar distribution along the
line of sight), this can be explained by the contribution of mirror
stars, which could naturally increase the optical depth towards
the galaxy bulge by factor 2 or so.

In dwarf galaxies mirror baryons may be less concentrated than in
spiral ones and may form relatively extended halos. Recent observational
data indeed suggest that the distribution of dark matter in dwarves
support the idea of dissipative dark matter \cite{dwarf}.

\vspace{8mm}
\noindent{\large \bf 3. Asymmetric reheating and asymmetry of
electroweak scales}
\vspace{4mm}

In view of the analysis of the previous section the
basic requirements to the model are
to provide a ground state in which both electroweak VEVs
are nonzero and different, to provide different
inflationary reheating in both worlds, and to suppress
all possible interactions which could bring two worlds into
thermal equilibrium with each other.

Let us consider a toy model involving two scalars
$\phi$, $\phi'$, analogues to our matter fields. Their
Lagrangian, invariant under the discrete transformation $P$:
$\phi \leftrightarrow \phi'$, has the form
\be{Lagr_phi}
{\cal V}_0(\phi,\phi')=(m_0^2 \phi^2 + h_0\phi^4) +
(m_0^2 \phi'^2 + h_0\phi'^4) +
a_0 \phi^2 \phi'^2
\ee
The simplest way to spontaneously break $P$-invariance
is to introduce a real $P$-odd scalar $\eta$ ($P:~\eta \to -\eta$),
with nonzero VEV \cite{CMP}.
For our model it is natural to assume that
$\eta$ also plays the role of inflaton.
Let us present the potential of $\eta$ in a generic form
${\cal V}(\eta)= \mu^4 {\cal P}(z)$
without specifying it in detail. Here
$z=\eta/M_{Pl}$ and ${\cal P}(z)$ is
a function that satisfies all necessary `inflationary' conditions,
with the Hubble parameter during inflation $H\sim \mu^2/M_{Pl}$
(for a review, see \cite{infl}).
The parameter $\mu$ is determined by the large scale density perturbations
(at scales which reenter the horizon in the matter dominated epoch):
$\delta\rho/\rho \simeq O(100) (\mu/M_{Pl})^2$,
so that the COBE result $\delta\rho/\rho \simeq 5\cdot 10^{-6}$
leads to $\mu \sim 10^{15.5}$ GeV. In the context of
generic inflation models this in turn
implies that the inflaton mass is
$m_\eta \sim \mu^2/M_{Pl} \sim 10^{12}$ GeV.

Discussing the couplings of $\eta$ to the matter fields,
we take into the account that in many
inflationary scenarios one deals with a large
($\sim M_{Pl}$) amplitude of the inflaton field and moreover,
in many cases its VEV is also $\sim M_{Pl}$.
Thus in general one cannot neglect higher order terms in $\eta$
and we formally sum up all them as follows
\be{toy_pot}
\mu^2 F(z)(\phi^2 + \phi'^2) +
\mu^2 \tilde{F}(z)(\phi^2 - \phi'^2) +
K(z) (\phi^4 + \phi'^4) + \tilde{K}(z) (\phi^4 - \phi'^4) +
A(z) \phi^2 \phi'^2 + \dots
\ee
where without loss of generality we take the dimensional parameter
as $\mu$, and absorb all uncertainties in the unknown factors
among which $F(z),~K(z)$, $A(z)$ are even functions of $z$
(vanishing at $z=0$), and
$\tilde{F}(z)$, $\tilde{K}(z)$ are odd functions.
Lacking understanding of the theory at Planckian energies,
we have no apriori information on the shape of these functions.
The effects of the possible kinetic-like terms
$G(z) [(\partial_\mu \phi)^2 + (\partial_\mu \phi')^2]$ and
$\tilde{G}(z) [(\partial_\mu \phi)^2 - (\partial_\mu \phi')^2]$
reduce to redefinition of the wavefunctions of $\phi,\phi'$
and for simplicity we do not consider them here.
As for the matter fields, we are interested only
in their small values, so that the possible higher order
terms in $\phi,\phi'$ can be neglected.

If $\eta$ develops a nonzero VEV $\langle \eta \rangle =\eta_0$, then
expanding in series of (small) deviations $\check{\eta}=\eta-\eta_0$
we see that neither the effective Lagrangian of
$\phi$ and $\phi'$, nor their interaction terms with
the inflaton field $\check{\eta}$ respect P-symmetry anymore:
\be{Lagr_phinew}
{\cal V}(\phi,\phi')=
(m^2 \phi^2 + h\phi^4) + (m'^2 \phi'^2 + h'\phi'^4) +  a \phi^2 \phi'^2
\ee
\be{Lagr_eta}
{\cal V}(\check{\eta};\phi,\phi')=(f\phi^2 + f'\phi'^2)\mu\check{\eta} +
(g \phi^2 + g'\phi'^2)\check{\eta}^2 +
(k\phi^4 + k'\phi'^4) \frac{\check{\eta}}{M_{Pl}} +
\dots
\ee
where in general the parameters are all different for
the primed and unprimed fields:
\beqn{parameters}
&& m^2 (m'^2)= m_0^2+\mu^2 (F \pm \tilde{F}), ~~~
h (h')= h_0 + (K \pm \tilde{K}), ~~~
a=a_0+A \nonumber \\
&& f (f')= \frac{\mu}{M_{Pl}} (F_z \pm \tilde{F}_z), ~~~
g (g')= \frac{\mu^2}{M_{Pl}^2} (F_{zz} \pm \tilde{F}_{zz}),~~~
k (k')= K_z \pm \tilde{K}_z
\eeqn
(here the values of the functions $F$, etc. and their derivatives
$F_z=dF/dz, ~F_{zz}=d^2F/dz^2$ etc. are taken at $z_0=\eta_0/M_{Pl}$).
Thus, in principle one can obtain both an asymmetric electroweak
breaking and an asymmetric postinflationary reheating.
%
For a certain range of parameters
both $\phi$ and $\phi'$ would have nonzero and different VEVs,
$v'\neq v$.
Since $P$ invariance is also broken in the inflaton
couplings (\ref{Lagr_eta}) the widthes $\Gamma$ and $\Gamma'$
of the decay of $\check{\eta}$ into $\phi$ and $\phi'$ particles
respectively are different. As a result, the reheating temperatures
in two worlds
$T_R\sim (\Gamma M_{Pl})^{1/2}$ and $T'_R\sim (\Gamma' M_{Pl})^{1/2}$
should also be different.\footnote{
It has been recently emphasized that in some cases
parametric resonance may amplify the particle production \cite{KLS}.
While in this case the numerical estimate of
the reheating temperature  is different, the fact
of asymmetric reheating will remain unchanged due to different coupling
of the inflaton to the two sectors. }

Let us first consider the case $\eta_0\sim M_{PL}$.
In this case a small size of the VEVs $v'$ and $v$ implies a strong
fine tuning. For example, for $v'\simeq 30 v$ being
about 5 TeV, the values of both $F$ and $\tilde{F}$ should be
$\sim 10^{-22}$. Furthermore, $F$ and $\tilde{F}$ should be also
fine tuned among each other with the accuracy
of $10^{-3}$ in order to get $v\sim 100$ GeV.

In order
to prevent mirror and usual particles from establishing
thermal equilibrium with each other, one has to suppress very much
the crossing term in (\ref{Lagr_phinew}):
$a < O(10)(m'/M_{Pl})^{1/2} \sim 10^{-7}$.
The same requirement puts an upper bound on
the values of $f,f'$. Indeed, for energies below the inflaton mass,
$m_\eta \sim \mu^2/M_{Pl}$, the first term in eq. (\ref{Lagr_eta})
mimics the contact term $\delta a \phi^2 \phi'^2$ with
$\delta a \sim ff'(\mu/m_\eta)^2$. Thus the above
"non-equilibrium" constraint still implies that $f,f' < 10^{-7}$.
This limit allows the reheating temperatures
$T_R$ and $T'_R$ to be as high as $10^{10}$ GeV.
However, once the functions $F(z)$ and $\tilde{F}(z)$ are small
$(\sim 10^{-22})$ at $z=z_0$, it would be unnatural if their
derivatives are substantially larger. In other words,
without additional fine tunings, these functions and
their derivatives should be
all very small $(\sim 10^{-22})$ for any $z$.
In this case two particle decays of inflaton into $\phi$ and $\phi'$
would lead to unacceptably small $T_R$ and $T'_R$.

As for $K$, $\tilde{K}$ and their derivatives $K_z$,
$\tilde{K}_z$, they are allowed to have values of order unity.
Indeed, the coupling constants $k,k'$ (as well as $h, h'$) can be
$\sim 1$ and different from each other.
Then 4-body decays of the inflaton lead
to different reheating temperatures $T_R$ and $T'_R$
of the typical `gravitational' size
$\sim 0.1 (m^3_\eta/M_{Pl})^{1/2} \sim 10^{7}$ GeV.

However, all these demand a very strong fine tuning between
parameters which is not natural. Moreover, in this case
everything becomes uncontrolable and arbitrary,
$P$ symmetry is actually broken already at the Planck scale
and in general it should be violated also in the Yukawa terms
due to the big $z$-induced corrections.
In other words, the mirror world becomes
a shadow world without any similarity to the visible one.

If $\eta$ gets VEV at some intermediate scale $\eta_0\ll M_{Pl}$,
for example in the chaotic inflation scenario with simple potential
${\cal V}(\eta)=h (\eta^2 - \eta_0^2)^2$ with $h\sim 10^{-15}$
(in the previous notations, this corresponds to
${\cal P}(z)=(z^2-z_0^2)^2$), then the unknown functions in
(\ref{toy_pot}) can be expanded in series of $z$.
The interaction terms (\ref{toy_pot}) then can be written as:
\be{small}
h_1 \eta^2 (\phi^2+\phi'^2) +
h_2 \eta_0\eta (\phi^2-\phi'^2) +
h_3 \frac{\eta^2}{M^2_{Pl}} (\phi^4+\phi'^4) +
h_4 \frac{\eta}{M_{Pl}}(\phi^4-\phi'^4) + \dots
\ee
One can achieve the (asymmetric) VEVs of
$\phi$ and $\phi$ fields in the TeV range
and acceptable reheating temperatures $T_R$ and $T'_R$
by choosing a proper range of $\eta_0$ and the constants in
(\ref{small}), however this still demands unnatural fine tunings.
In the next section we present more appealing supersymmetric
models for asymmetric inflationary reheating.

\vspace{7mm}
\noindent{\large \bf 4. SUSY models for asymmetric
inflationary reheating}
\vspace{4mm}

In this section we outline a supersymmetric model for
chaotic inflation which leads to asymmetric reheating of the
normal and the mirror sectors and connects this asymmetry
to the different electroweak breakings in the two sectors.
We assume the minimal supersymmetric standard
model (MSSM) for both sectors denoting the Higgs superfields in
the visible and mirror sectors respectively as $H_u,~H_d$
and $H'_u,~H'_d$, and  introduce also a gauge singlet
superfield $\eta$.
We also assume that $P$ parity is realized as a discrete $R$ symmetry
under which $H_u,H_d \to H'_u,H'_d$, $\eta \to -\eta$ and
$W \to -W$, and consider the following superpotential:
\be{sup}
W= \frac{\lambda}{3}\, \eta^3 + \lambda_1\eta(H_uH_d + H'_uH'_d)
-\lambda_2\frac{\eta^2}{M_{Pl}}(H_uH_d - H'_uH'_d)
\ee
We do not include the terms linear in $\eta$ and bare mass terms
of the doublets, assuming that all mass terms arise
purely from the soft SUSY breaking scale $m\sim $ few TeV.
Then the Higgs potential of the scalar $\eta$ has the form
\be{pot_eta}
{\cal V}(\eta)= \lambda^2 |\eta|^4 + Am \lambda (\eta^3 + \eta^{\ast 3})
+ m^2 |\eta|^2
\ee
while the part of the potential involving the Higgs doublets is
\beqn{pot}
{\cal V}(\eta;H,H')
&&
=\left|\lambda_1-2\lambda_2{{\eta}\over{M}}\right|^2|H_u H_d|^2 +
\left|\lambda_1+2\lambda_2{{\eta}\over{M}}\right|^2|H'_u H'_d|^2
\nonumber \\
&&
+\left[\left(\lambda_1^2- 2\lambda_1\lambda_2\frac{\eta-\eta^\ast}{M} +
4\lambda_2^2\frac{|\eta|^2}{M^2}\right)(H_uH_d)(H'_uH'_d)^\ast +
{\rm h.c.}\right]   \nonumber \\
&&
+ \left| \lambda_1 -\lambda_2{{\eta}\over{M}}\right|^2
|\eta|^2 (|H_u|^2+|H_d|^2)+
\left| \lambda_1 +\lambda_2{{\eta}\over{M}}\right|^2
|\eta|^2 (|H'_u|^2+|H'_d|^2)  \nonumber \\
&&
+ \left[\lambda \left(\lambda_1 -2\lambda_2\frac{\eta^\ast}{M}\right)
\eta^2 (H_uH_d)^\ast +
\lambda \left(\lambda_1 +2\lambda_2\frac{\eta^\ast}{M}\right)
\eta^2 (H'_uH'_d)^\ast + {\rm h.c.}\right] \nonumber \\
&&
+ m\left[\left(B\lambda_1 -C\lambda_2\frac{\eta}{M}\right)\eta H_uH_d  +
\left(B\lambda_1+C\lambda_2\frac{\eta}{M}\right)\eta H'_uH'_d +
{\rm h.c.} \right]    \nonumber \\
&&
+ \kappa m^2(|H_u|^2+|H_d|^2+|H'_u|^2+|H'_d|^2) ~+ ~{\rm D-terms}
\eeqn
where $A,B,C,\kappa$ are the $O(1)$ constants determined
by the SUSY breaking hidden sector.

Clearly, the potential (\ref{pot_eta}) is suitable for chaotic
inflation \cite{infl}
and in order to obtain acceptable density fluctuations we have to
assume that $\lambda^2 \sim 10^{-15}$, or $\lambda \sim 10^{-7.5}$.
Hence, the scalar part of the $\eta$ superfield can play a role of
the inflaton and its interactions with the Higgs doublets determine
the nature of the postinflationary reheating.\footnote{
Strictly speaking, the potential (\ref{pot_eta}) occurs in the case
of the global supersymmetry. In the minimal supergravity scheme the
standard factor $\exp (8\pi|\eta|^2/M^2_{Pl})$ in the potential would
spoil the slow roll condition for $\eta\geq M_{Pl}$. However, one can
appeal to the no-scale supergravity scheme with a K\"ahler potential
suggested in \cite{GL}, in which case the relevant part of the theory
works as in a global SUSY case. }

If $A>1$, then  the absolute
minimum of (\ref{pot_eta}) is achieved for nonzero $\eta$:
$\eta_0\sim Am/\lambda \sim 10^{12}$ GeV, which spontaneously
breaks $P$-invariance.\footnote{
Notice that there is no problem of domain walls,
since the discrete symmetry $\eta \to -\eta$ is explicitly
violated by the trilinear soft term in (\ref{pot_eta}).
In addition, an accidental discrete symmetry
$\eta \to \exp(\frac{2\pi}{3}{\rm i})\eta$
of the potential (\ref{pot_eta}) is also explicitly
violated by the third term in the superpotential (\ref{sup}). }
This will immediately induce asymmetry in the
electroweak scales. Indeed,
substituting $\eta \to \eta_0$ in the potential (\ref{pot})
we see that the mass terms of the Higgs doublets
become different in the visible and mirror sectors.
(Notice that $\mu$-terms also are induced,
which are asymmetric too.)
Taking $\lambda_1 \sim 10^{-7}$ (as well as $\lambda$)
and $\lambda_2 \sim 1$
(so that $\lambda_2\eta_0/M_{Pl}\sim 10^{-7}$ too),
one gets mass terms of $H$ and $H'$ fields
in the TeV range but different from each other.
Extrapolating this via renormalization group to lower energies we see
that the VEVs
$\langle H_d \rangle = v_1$ and $\langle H_u \rangle = v_2$
are different from the VEVs of the mirror doublets
$\langle H'_d \rangle = v'_1$ and $\langle H'_u \rangle = v'_2$.
(In order to obtain the standard electroweak scale
$v=(v_1^2+v_2^2)^{1/2}$ by an order of magnitude smaller than the
mirror scale $v'=({v'}_1^2+{v'}_2^2)^{1/2}\sim$ few TeV,
one has to allow $\sim 0.1$ fine tuning, which seems
to be reasonable.)
Moreover, in general the `up-down' ratios $\tan\beta=v_2/v_1$
and $\tan \beta'=v'_2/v'_1$ are also different in two sectors.
This can alter the content of the mirror
baryonic matter discussed in Section 2. In particular, for
$\tan\beta'>2\tan\beta$ the only stable nucleon in the mirror
world would be the mirror neutron.

The second and the third terms in eq. (\ref{sup}) combine to give
different decay rates for
the inflaton field $\eta$ into visible and mirror particles,
which leads to different reheating temperatures, $T'_R\neq T_R$.
Since its mass $m_\eta\sim m \sim $ few TeV, the
decay widthes into different states are approximately
$\sim \lambda^2 m$,  which leads to the reheating
temperatures $T_R,T'_R$ around a few TeV.

The low reheating temperature
excludes many possible mechanisms of baryogenesis but still the
electroweak one remains. We have estimated
$T_R$ to be of the order of the mirror electroweak scale $v'$,
but much larger than the standard electroweak scale $v$. Hence,
it is likely that after inflationary reheating the
mirror universe was already in the phase of the broken electroweak
symmetry,\footnote{In this case the initial number of species in
the mirror world $g'_*$ should be smaller than that in the visible
world and the estimates of the Section 2 should be correspondingly
changed.}
so that the mirror BA might be very small (though the case
of a large mirror BA, even larger than the normal one,
is by no means excluded).
On the other hand, the visible world reheats in unbroken
phase. Then in our SUSY model the electroweak phase transition can
be easily first order, and the observed BA can be produced due to
the (supersymmetric) electroweak baryogenesis mechanism \cite{krs}.
Moreover, due to supercooling and
additional entropy production at the first order phase transition
one can suppress the abundance of the mirror particles even if
initial reheating temperatures were the same.

In the spirit of this observation, one could in fact replace the
last (non-renormalizable) term in (\ref{sup}) by the mass term
$\mu (H_uH_d - H'_uH'_d)$ with $\mu \sim m$.
This term in combination with the second term in (\ref{sup})
will still cause asymmetry between the VEVs $v$ and $v'$.
However, in this case inflaton
couples to $H$ and $H'$ fields in a symmetric way, and the
temperature difference between the visible and mirror worlds
at the BBN epoch ($T'<T$) can arise purely from the difference
in the electroweak scales ($v'>v$), due to the
postinflationary `miniinflation' during the (possible) first order
phase transition in the visible universe as well as due to the
different phase space factors in the inflaton decays into the usual
and mirror particles (since the masses of the latter are closer
to the inflaton mass $m_\eta\sim $ few TeV).

Below we present another model which does not have the usual
fine tuning problems
and can also provide much larger reheating temperatures.
Let us consider the supersymmetric $SU(6)\times SU(6)'$ model which
could emerge in the context of the $E_8\times E'_8$ string theory.
The SUSY GUT $SU(6)$ \cite{su6} has an advantage that it
explains the doublet-triplet splitting without fine tuning.
The Higgs sector consists of the superfields $\Sigma$ and $H+\bar H$
respectively in adjoint 35 and fundamental $6+\bar 6$ representations.
The Higgs doublets appear to be light as
the pseudo-Goldstone modes  of the spontaneously broken
accidental global symmetry $SU(6)_\Sigma \times SU(6)_H$ which
arises if the crossing terms like $H\Sigma \bar H$
are suppressed in the superpotential.
At the scale $V_H\sim 10^{17}$ GeV, $H,\bar H$ break
the local $SU(6)$ symmetry down to $SU(5)$ which is then broken
down to $SU(3)\times SU(2)\times U(1)$ by $\Sigma$
at the scale $V_\Sigma\simeq 10^{16}$ GeV.
In this case the Higgs doublets
$H_{u,d}$ dominantly come from the doublet fragments of $\Sigma$
while in $H,\bar H$ they are contained with the weight
$V_\Sigma/V_H$, and the observed hierarchy of fermion masses
can be naturally explained in terms of the small parameter
$V_\Sigma/V_H\sim 0.1$ \cite{su6}. Needless to say that the mirror
group $SU(6)'$ is assumed to be exactly the same.

For inflation we take
the model similar to the one suggested in \cite{DSS}. It
includes the $P$-even ($S$) and $P$-odd ($\eta$) singlet superfields.
We assume the following superpotential:
\be{DSS}
W=k S (\eta^2 - \mu^2) + a S \frac{\eta}{M}(\bar{H}H - \bar{H}'H')
+ b S \frac{\eta^2}{M^2} (\Sigma^2 + \Sigma'^2) + \dots
\ee
where $\mu\sim 10^{15.5}$ GeV and $k,a,b\sim 10^{-2}$.
The vacuum state is $\langle \eta \rangle = \mu$ and
$\langle S \rangle = 0$. The tree level potential
of inflaton $S$ appears to be flat for large values of the field
$S > \mu$. However radiative effects remove the degeneracy
and provide necessary ``inflationary" profile \cite{DSS}.
The superpotential (\ref{DSS}) has an advantage that the slow
roll conditions are satisfied for the values of the inflaton field
smaller than $M_{Pl}$ ($S\sim 10^{17}$ GeV).
Hence the Planck scale corrections are irrelevant and the
model can be safely incorporated into the minimal supergravity scheme.
The last two terms in (\ref{DSS}) combine to give
different decay rates of the inflaton into usual and mirror Higgs
doublets, so that the reheating temperatures $T_R$ and $T'_R$ in
two worlds are different and have the typical magnitude
$\sim 10^8$ GeV. It is important to stress that the
coupling constants ($k,a,b$ etc.) need not be taken extremely small
since the large VEV of $\eta$ does not induce the large
mass terms for the Higgs doublets.
The asymmetry of the electroweak scales in two sectors can
naturally emerge as a result of the $P$ parity violation in the
soft SUSY breaking terms.


\vspace{8mm}
\noindent
{\large \bf 5. Conclusions }
\vspace{4mm}

   To summarize, we have discussed cosmological implications of
the idea that there is a parallel mirror universe with identical
gauge and matter content to the one we inhabit. Consistency with
the big bang nucleosynthesis requires that the energy density of the
mirror particles should be suppressed with respect to the normal
ones. This can be either achieved by a weaker reheating of the
mirror world after inflation or by diluting mirror particles
by the entropy production in the first order phase transition
in the usual world. The latter could be achieved if the usual
electroweak scale is below the mirror one.
We have given explicit examples showing how this idea can
be realized in realistic models and discussed
how this asymmetry may be connected to
the different values of the electroweak symmetry breaking scales
in the two sectors. At this point our approach drastically differs
from the mirror universe models with exact $P$ parity
\cite{FV,mirror}.
We have also discussed other cosmological aspects of the mirror universe.
In particular, we argue that the mirror baryonic matter is
likely to consist only of hydrogen and no heavier
nuclei, thus there should be
no mirror stars with a thermonuclear active core. There might be
cold mirror compact bodies around which could be accessible to microlensing
searches. Mirror baryons might form early black holes explaining quasar
formation and active galactic nuclei. Mirror baryonic dark matter should
be rather different from the usual dark matter because it is dissipative
and can possess antibiasing features.

\vspace{7mm}
\noindent{\bf Acknowledgements.}
Z.G.B. thanks R. Ansari, D. Caldwell, G. Dvali,
G. Fiorentini and S.S. Gershtein,
and R.N.M. thanks P. Ferreira, H. Murayama and S. Nussinov
for useful discussions.
R.N.M. would like to thank the theory
group at Lawrence Berkeley Laboratory, and J. Silk and the Center for
Particle Astrophysics at the university of California, Berkeley for
kind hospitality during the time when this work was completed.
The work of R.N.M is supported by the National
Science Foundation under grant No. PHY9421385 and a Distinguished
Faculty Research Award by the University of Maryland.
The work of A.D.D is
supported by DGICYT under grants PBS2-0084 and SUB94-0089.


\begin{thebibliography}{99}


\bibitem{Witten}
M. Green, J. Schwarz and E. Witten, {\it Superstring Theories},
Cambridge University press (1989).

\bibitem{BM} Z.G. Berezhiani and R.N. Mohapatra,
hep-ph/9505385, Phys. Rev. D 52  (in press).

\bibitem{FV} R. Foot and R. Volkas, hep-ph/9505359;
see also Z. Silagadze, hep-ph/9503481.

\bibitem{mirror}
T.D. Lee and C.N. Yang, Phys. Rev. 104 (1956) 254;
A. Salam, N. Cim. 5 (1957) 299;
Y. Kobzarev, L. Okun and I. Pomeranchuk, Yad. Fiz. 3 (1966) 1154;
L.B. Okun, ZhETF 79 (1980) 694;
S.I. Blinnikov and M.Yu. Khlopov, Astron. Zh. 60 (1983) 632;
B. Holdom, Phys. Lett. B 166 (1985) 196;
S. Glashow, {\em ibid.} 167 (1986) 35;
E. Carlson and S. Glashow, {\em ibid.} 193 (1987) 168;
R. Foot, H. Lew and R. Volkas, {\em ibid.} 272 (1991) 67;
Mod. Phys. Lett. A7 (1992) 2567;
M. Khlopov et al., Astron. Zh. 68 (1991) 42.

\bibitem{calmoh}
D. Caldwell and R.N. Mohapatra, Phys. Rev. D 48 (1993) 3259;
J. Peltoniemi and J.W.F. Valle, Nucl. Phys. B 406 (1993) 409.

\bibitem{sch} V.F. Schwartzmann, JEPT Lett. 9 (1969) 184;
for the recent analysis see K.A. Olive and G. Steigman Ap. J. Suppl.
97 (1995) 49; C.J. Copi, D.W. Schramm and M.S. Turner,
Science 267 (1995) 192,  and references therein.

\bibitem{BEG} R. Barbieri, J. Ellis and M.K. Gaillard, Phys. Lett.
B 90 (1980) 249.

\bibitem{ABS} E. Akhmedov, Z. Berezhiani and G. Senjanovi\'c,
Phys. Rev. Lett. 69 (1992) 3013.

\bibitem{MSW} S. Mikheyev and A. Smirnov, Yad. Fiz. 42 (1985) 1441;
L. Wolfenstein, Phys. Rev. D 17 (1978) 2369.
For the recent status see P. Krastev and A. Smirnov, Phys. Lett.
B 338 (1994) 282; V. Berezinsky, G. Fiorentini and M. Lissia,
Phys. Lett. B 341 (1994) 38; N. Hata and P. Langacker,
Phys. Rev. D 50 (1994) 632.

\bibitem{justso} V. Gribov and B. Pontecorvo,
Phys. Lett. 28 (1967) 493;
J. Bahcall and S. Frautschi, Phys. Lett. 29B (1969) 623;
V. Barger, R. Phillips and K. Whisnant,
Phys. Rev. D 24 (1981) 538;
for the recent status see
P. Krastev and S. Petkov, Phys. Rev. Lett. 72 (1994) 1690
and Preprint SISSA 09/95/EP;
F. Calabresu et al., hep-ph/9507352;
Z.G. Berezhiani and A. Rossi, Phys. Rev. D 51 (1995) 5229 and
hep-ph/9507393.

\bibitem{bd} R. Barbieri and A. Dolgov, Phys. Lett. B 237 (1990)
440; Nucl. Phys. B 349 (1991) 743;
K. Kainulainen, Phys. Lett. B 244 (1990) 191;
K. Enqvist, K. Kainulainen and M. Thomson, Nucl. Phys. B 373 (1992)
498.

\bibitem{KST} E.W. Kolb, D. Seckel and M.S. Turner, Nature 514
(1985) 415; see also H.M. Hodges, Phys. Rev. D 47 (1993) 456.


\bibitem{TG} S. Tremaine and J. Gunn, Phys. Rev. Lett. 42 (1979) 407.

\bibitem{WDM} S. Dodelson and L. Widrow, Phys. Rev. Lett. 72
(1994) 17;
R.A. Malaney, G.D. Starkman and L. Widrow, Report No.
CITA-95-9 (April 1995).

\bibitem{HDM2} J.R. Primack et al.,
Phys. Rev. Lett. 74 (1995) 2160 and references therein.

\bibitem{Ansari}
R. Ansari, Nucl. Phys. B (Proc. Suppl.) 43 (1995) 108.

\bibitem{DeRuj}
A. De Rujula et al., Astron. and astrophys. 254 (1992) 99.

\bibitem{OGLE} A. Udalski et al., Ap. J. 426 (1994) L69;
Ch. Alcock et al., astro-ph/9506016.

\bibitem{dwarf}
B. Moore, Nature 370 (1994) 629; R.A. Flores and
J.R. Primack, Ap. J. 427 (1994); A. Burkert, astro-ph/9504041.

\bibitem{CMP} D. Chang, R.N. Mohapatra, M.K. Parida,
 Phys. Rev. D 30 (1984) 1052.

\bibitem{infl} A. Linde, {\em Particle Physics and Inflationary
Cosmology}, Harwood, Switzerland, 1990;
K.A. Olive, Phys. Rep. 190 (1990) 307;
A. Liddle and D. Lyth, {\em ibid.} 231 (1993) 1.

\bibitem{KLS} L. Kofman, A.D. Linde and A.A. Starobinsky,
Phys. Rev. Lett. 73 (1994) 3195;
Y. Shtanov, J. Traschen and R. Brandenberger, Phys. Rev. D 51
(1995) 5438;
D. Boyanovsky et al., hep-ph/9507414; for the earlier papers see
A.D. Dolgov and D.K. Kirilova, Sov. J. Nucl. Phys. 51 (1990) 273;
J. Traschen and R. Brandenberger, Phys. Rev. D 42 (1990) 2491.

\bibitem{GL} A.S. Goncharov and A.D. Linde,
Clas. Quant. Grav. 1 (1984) L75.

\bibitem{krs} V.A. Kuzmin, V.A. Rubakov, and M.E. Shaposhnikov,
Phys. Lett. B155 (1985) 36; for the review see
A.D. Dolgov, Phys. Rep. 222 (1992) 309; A.G. Kohen, D.B. Caplan
and A.E. Nelson, Annu. Rev. Nucl. Part. Sci. 43 (1993) 27.


\bibitem{su6} Z.G. Berezhiani and G.R. Dvali,
Sov. Phys. -- Lebedev Inst. reports 5 (1989) 55; \\
R. Barbieri, G. Dvali and M. Moretti, Phys. Lett. B 312 (1993) 137;
R. Barbieri et al., Nucl. Phys. B 432 (1994) 49; ~
Z.G. Berezhiani, Phys. Lett. B 355 (1995) 481;
Z.G. Berezhiani, C. Csaki and L. Randall, Nucl. Phys. B 444 (1995) 61.

\bibitem{DSS} G. Dvali, Q. Shafi and R. Schaefer, Phys. Rev. Lett.
73 (1994) 1886.

\end{thebibliography}
\end{document}